\documentclass[aps,prb,twocolumn,showpacs,groupedaddress]{revtex4}

\usepackage{graphicx}

\newcommand{\etal}{{\it et~al.}}

\begin{document}

\preprint{}

\title{Strong electronic correlation and strain effects at the interfaces between polar and nonpolar complex oxides}

\author{A. Annadi$^{1,2}$}

\author{A. Putra$^{1,3}$}

\author{Z. Q. Liu$^{1,2}$}

\author{X. Wang$^{1,2}$}

\author{K. Gopinadhan$^{1,4}$}

\author{Z. Huang$^{1}$}

\author{S. Dhar$^{1,4}$}

\author{T. Venkatesan$^{1,2,4}$}

\author{Ariando$^{1,2}$}

\altaffiliation[Email: ]{ariando@nus.edu.sg}

\affiliation{$^1$NUSNNI-Nanocore, National University of Singapore, 117411 Singapore}

\affiliation{$^2$Department of Physics, National University of Singapore, 117542 Singapore}

\affiliation{$^3$Department of Engineering Science Programme, National University of Singapore, 117576 Singapore}

\affiliation{$^4$Department of Electrical and Computer Engineering, National University of Singapore, 117576 Singapore}

\begin{abstract}
The interface between the polar LaAlO$_3$ and nonpolar SrTiO$_3$ layers has been shown to exhibit various electronic and magnetic phases such as two dimensional electron gas, superconductivity, magnetism and electronic phase separation. These rich phases are expected due to the strong interplay between charge, spin and orbital degree of freedom at the interface between these complex oxides, leading to the electronic reconstruction in this system. However, until now all of these new properties have been studied extensively based on the interfaces which involve a polar LaAlO$_3$ layer. To investigate the role of the A and B cationic sites of the ABO$_3$ polar layer, here we study various combinations of polar/nonpolar oxide (NdAlO$_3$/SrTiO$_3$, PrAlO$_3$/SrTiO$_3$ and NdGaO$_3$/SrTiO$_3$) interfaces which are similar in nature to LaAlO$_3$/SrTiO$_3$ interface. Our results show that all of these new interfaces can also produce 2DEG at their interfaces, supporting the idea that the electronic reconstruction is the driving mechanism for the creation of the 2DEG at these oxide interfaces. Furthermore, the electrical properties of these interfaces are shown to be strongly governed by the interface strain and strong correlation effects provided by the polar layers. Our observations may provide a novel approach to further tune the properties of the 2DEG at the selected polar/nonpolar oxide interfaces.
\end{abstract}



\maketitle
\section{Introduction}

Heterostructures constructed from polar/nonpolar oxides, especially the LaAlO$_3$/SrTiO$_3$ interface, have been a topic of research in recent years in which along with 2DEG, novel properties have been reported [1-5]. Further, it has been demonstrated that the properties at the LaAlO$_3$/SrTiO$_3$ interface can be manipulated in many ways such as electric field effect [2], AFM charge writing [6] and adsorbates capping [7], showing the ability to tune the interface by external parameters. Moreover, in LaAlO$_3$/SrTiO$_3$ heterostructure the combined effects of lattice mismatch ($\sim2.3\%$) between the LaAlO$_3$ and SrTiO$_3$ and electronic effects provided by polar LaAlO$_3$ at the interface play a strong role in controlling the interface properties. Recently, C. W. Bark \etal [8] demonstrated strain effects on properties of 2DEG at the LaAlO$_3$/SrTiO$_3$ interface by growing the heterostructure on different substrates. In addition to this, electrostriction and electromechanical response were demonstrated at the LaAlO$_3$/SrTiO$_3$ heterostrures [9,10]. Recently, a similar 2DEG is also shown to exist at the LaGaO$_3$/SrTiO$_3$ interface [11]. In all above observations, the polar nature of the over layer is critical. The significance of the polar over layer can be further explored by investigating various new combinations of polar/nonpolar oxide heterostructures. In this scenario, a choice of variety in chemical nature and lattice structure of polar layers enable us to tune the interface properties. Moreover this approach may help in revealing key issues such as driving mechanism of the formation of 2DEG which is widely believed to be originating from polarization catastrophe [12] due to this polar/nonpolar nature of the interface and the localization of carriers at this interface due to the lifting of the degeneracy of the Ti $d$ states. In this article, we investigate various combinations of polar/nonpolar oxide interfaces (Fig. 1(a)) REBO$_3$/SrTiO$_3$ (100) (RE=La, Pr, Nd, B=Al, Ga) and study their structural and electrical properties. The choice of polar oxides is made in such a way that all of them are similar to the LaAlO$_3$ in nature, having (AO)$^{+1}$, (BO$_2$)$^{-1}$ polar charge layers alternatively along (100) direction.

\section{Experimental}

\begin{figure}
\includegraphics[width=3.4in]{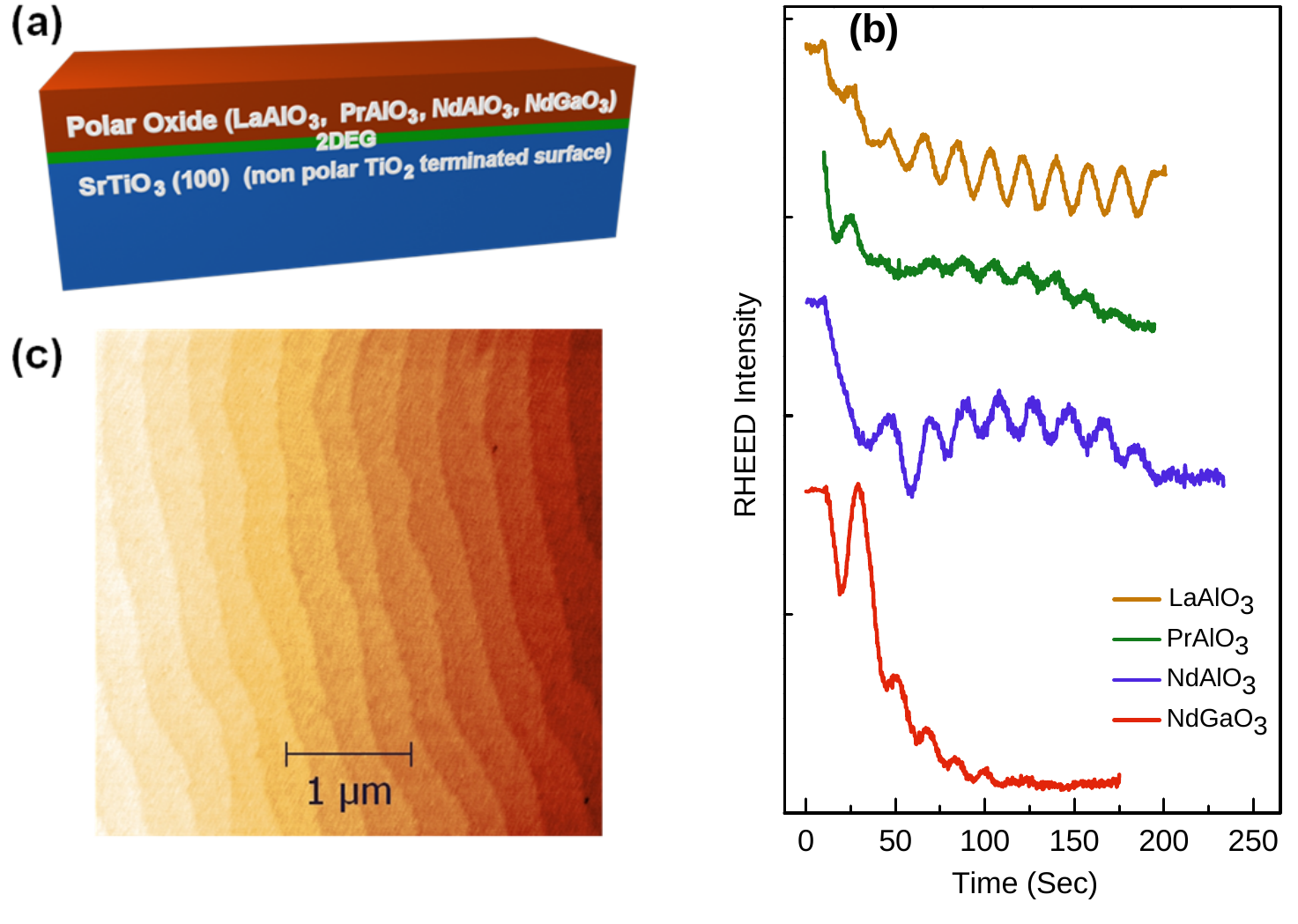}
\caption{\label{fig1} (a) Schematic representation of polar/nonpolar REBO$_3$/SrTiO$_3$ (RE-Rear Earth element) heterostructure. (b) Typical RHEED oscillations obtained for the growth of different polar layers on top of TiO$_2$ terminated SrTiO$_3$ substrate. (c) Atomic force microscopy (AFM) image of 10 uc NdAlO$_3$/SrTiO$_3$ sample showing an atomically smooth surface.}
\end{figure}

Samples were prepared by pulsed laser deposition, ablating REBO$_3$ targets onto TiO$_2$ terminated SrTiO$_3$ (100) substrates. Atomically flat TiO$_2$ terminated SrTiO$_3$ (100)  substrates  were obtained by employing well established conditions; buffer HF treatment for 30 sec followed by thermal annealing at $950 ^\circ$C for 2 hours [13,14]. Samples were prepared in a range of oxygen partial pressure ($P_{O2}$) of $1\times10^{-3}$ to $1\times10^{-5}$ Torr at $800 ^\circ$C. The laser (248 nm) energy density was 1.4 J/cm$^2$ and repetition rate was 1 Hz. During deposition, the film growth and number of unit cells were monitored by in-situ reflection high energy electron diffraction (RHEED). Samples with thicknesses ranging from 1-14 unit cells (uc) were grown for NdAlO$_3$ while a fixed 10 uc was used for all other polar layers for comparison. Epitaxial growth of films on SrTiO$_3$ substrates were further confirmed by high resolution X-ray diffraction (HRXRD). Electrical contacts were made with an Al wire-bonding directly to the interface. Sheet resistance and Hall Effect measurements were carried out to extract carrier density, mobility and electrical properties. Figure 1(b) shows RHEED intensity patterns obtained during the growth of the samples, clearly showing the layer by layer growth mode for various polar oxide layers on SrTiO$_3$. Figure 1(c) shows AFM topography image after the growth of 10 uc NdAlO$_3$ on SrTiO$_3$ with preserved steps further confirms the layer by layer growth. Even though all these polar oxides offer substantial lattice mismatch with SrTiO$_3$ substrate, we have successfully grown them by optimizing deposition conditions.

\section{Results and Discussion}

We first discuss the case of NdAlO$_3$/SrTiO$_3$ interface. Figure 2(a) shows the dependence of sheet resistance, $R_s$, with temperature for the NdAlO$_3$/SrTiO$_3$ interface prepared in the $P_{O2}$ range of $1\times10^{-3}-1\times10^{-5}$ Torr. While all samples clearly exhibit a conducting property at their interface, the high $P_{O2}$ grown samples show strong upturns in $R_s$ at lower temperatures. Interestingly, even the sample grown at $P_{O2}$ of $1\times10^{-4}$ Torr shows an upturn. It is noted here that in the case of LaAlO$_3$/SrTiO$_3$ (100) interfaces such an upturn emerges only for higher $P_{O2}$ ($\geq10^{-3}$ Torr) and ``thicker" samples ($\geq15$ uc) [4,15]. However, the sample grown at $P_{O2}$ $1\times10^{-5}$ Torr showed higher conductivity which can be understood in terms of the presence of oxygen vacancies at the interface. Figures 2(b) and 2(c) show temperature dependence of carrier density, $n_s$, and mobility, $\mu$, for the NdAlO$_3$/SrTiO$_3$ samples, respectively. For high $P_{O2}$ samples, $n_s$ (300 K) is of the order of $4-5\times10^{13}$ cm$^{-2}$. Whereas for sample grown at $1\times10^{-5}$ Torr, $n_s$ (300 K) is of the order of $10^{16}$ cm$^{-2}$, further confirming the presence of oxygen vacancies which is a common observation for the interfaces grown at lower $P_{O2}$ [16,17]. A large carrier freeze-out towards low temperatures is observed in high $P_{O2}$ samples. This kind of behaviour was previously observed in the cases of LaAlO$_3$/SrTiO$_3$ [18]. For all the samples, $\mu$ (300 K) is of the order of 1-8 cm$^2$V$^{-1}$s$^{-1}$ comparable to those reported for the LaAlO$_3$/SrTiO$_3$ (100) interfaces [15,19]. The $\mu$ follows $T^{-2}$-like dependence in the temperature range of 300-100 K, typical behaviour of Fermi liquid. However, a significant difference in $\mu$ is observed at low temperatures between samples grown at high and low $P_{O2}$. While an abrupt drop in $\mu$ towards low temperatures is observed (which generally occurs when strong localization/magnetic scattering effects present in the system) for high $P_{O2}$ samples, an increase in $\mu$ throughout the temperature range down to 2 K is observed for the sample grown at $1\times10^{-5}$ Torr.

\begin{figure}
\includegraphics[width=3.4in]{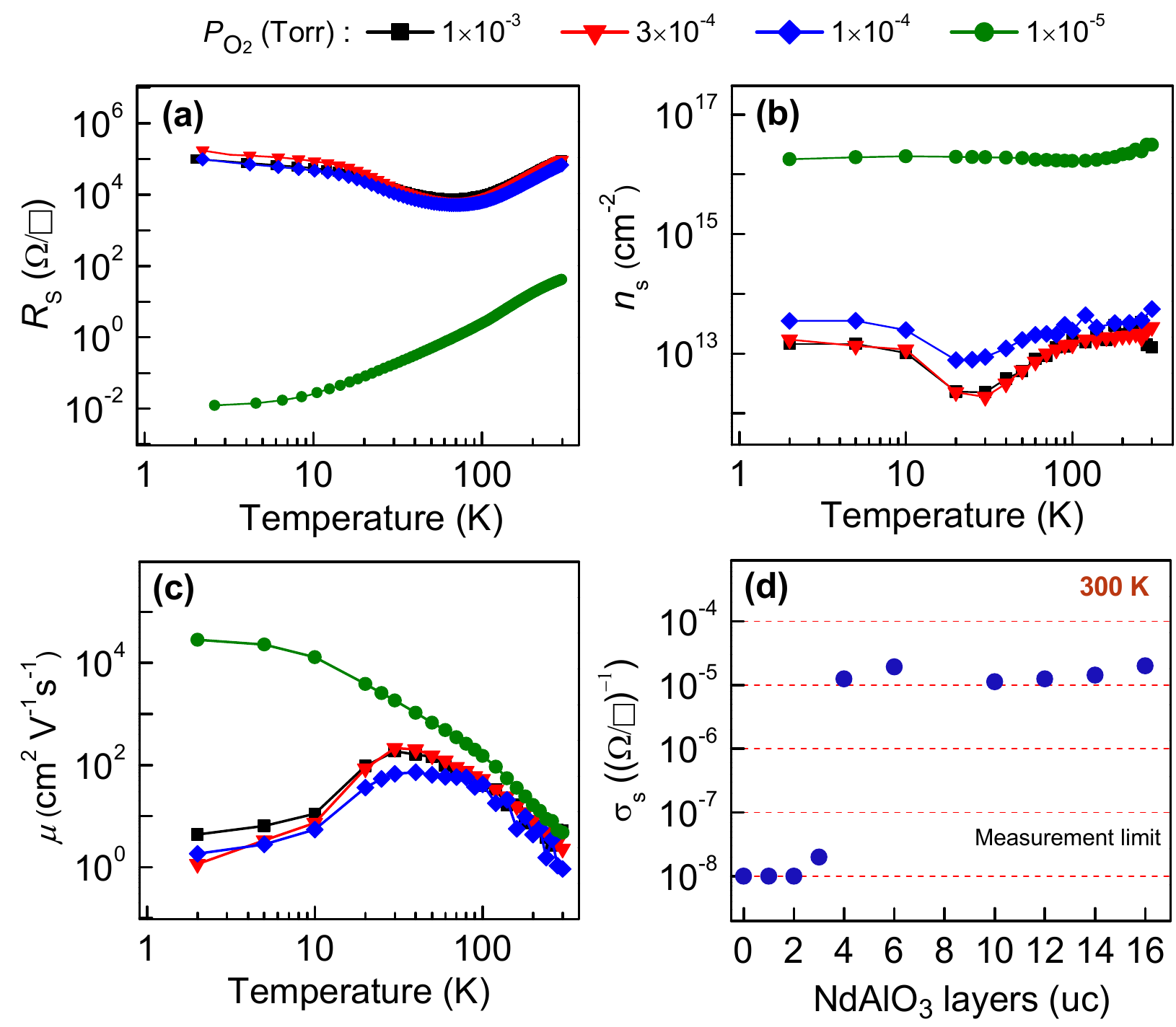}
\caption{\label{fig2} (a) Temperature dependence of the sheet resistance, $R_s(T)$, of the NdAlO$_3$/SrTiO$_3$ interfaces, for different oxygen partial pressures ($P_{O2}$) during growth. Temperature dependence of (b) carrier density $n_s$ and (c) Hall mobility $\mu$ of the NdAlO$_3$/SrTiO$_3$ interfaces. (d) Dependence of the sheet conductivity on the number of NdAlO$_3$ unit cells for the NdAlO$_3$/SrTiO$_3$ interfaces.}
\end{figure}

One of the characteristic features of the 2DEG at the LaAlO$_3$/SrTiO$_3$ system is thickness dependence of the insulator-metal transition with a critical thickness of $\sim4$ uc of the LaAlO$_3$ [2]. Sheet conductivity for the NdAlO$_3$/SrTiO$_3$ interfaces as a function of the number of unit cells of NdAlO$_3$ is depicted in Fig. 2(d), showing a clear transition from insulating to metallic interface at a thickness around 4 uc, with a change in the conductivity of more than three orders of magnitude across the insulator-metal transition. The similarity in the insulator-metal transition to the LaAlO$_3$/SrTiO$_3$ system confirms the formation of 2DEG at the NdAlO$_3$/SrTiO$_3$ interface and supports the idea that the electronic reconstruction could also be the driving mechanism in these interfaces. However, substantial differences observed in temperature dependence of the $n_s$ and $\mu$  indicate the effect of the different polar layers.

\begin{figure}
\includegraphics[width=3.4in]{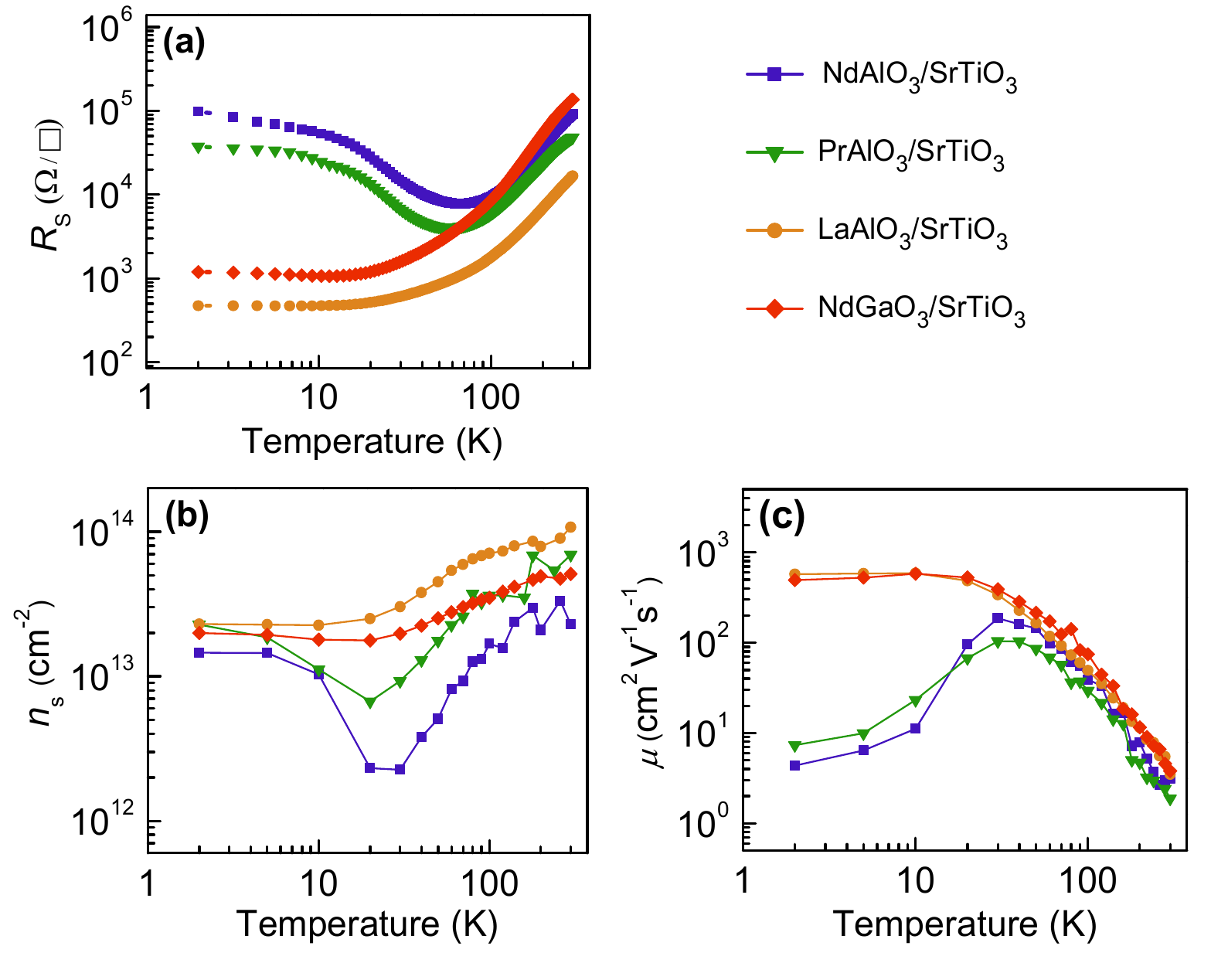}
\caption{\label{fig3} Comparison of transport properties of different polar/nonpolar (REBO$_3$/SrTiO$_3$ RE=La, Pr, Nd, B=Al, Ga) oxide interfaces. (a) Temperature dependence of $R_s$ of the interfaces. Temperature dependence of (b) $n_s$ and (c) $\mu$ of the corresponding interfaces.}
\end{figure}

To further elucidate the polar layer effects a comparison study was carried out and samples with 10 uc LaAlO$_3$, PrAlO$_3$ and NdGaO$_3$ as the polar over layers were grown at $P_{O2}=1\times10^{-3}$ Torr. Figure 3(a) shows the $R_s$ variation with temperature for these various polar/nonpolar combinations. Similar to the case of NdAlO$_3$/SrTiO$_3$, PrAlO$_3$/SrTiO$_3$ interfaces also show upturn in $R_s$ at low temperatures. On the other hand, interestingly, LaAlO$_3$/SrTiO$_3$ and NdGaO$_3$/SrTiO$_3$ interfaces show a typical metallic behaviour without appreciable upturns in $R_s$. Further, their $R_s$ at 300 K appears to have a strong dependence on the RE cation of the polar layer, where $R_s$ is larger ($\sim{90}$ k$\Omega$) for NdAlO$_3$ compared to that of LaAlO$_3$ ($\sim{20}$ k$\Omega$). Figure 3(b) shows the $n_s$ variation with temperature; LaAlO$_3$/SrTiO$_3$ and NdGaO$_3$/SrTiO$_3$ interfaces show minute carrier freeze-out towards low temperatures compared to NdAlO$_3$/SrTiO$_3$ and PrAlO$_3$/SrTiO$_3$, indicating that the interfaces with LaAlO$_3$ and NdGaO$_3$ have less contribution from activated carriers. However, a carrier recovery  below 15 K is observed for the NdAlO$_3$/SrTiO$_3$ and PrAlO$_3$/SrTiO$_3$ interfaces, which is possibly due to the difference transport mechanism in this low temperature localized regime (further investigation is needed to understand this unusual behaviour). The most remarkable effects are seen in temperature dependence of mobility which is shown in Fig. 3(c). For NdAlO$_3$/SrTiO$_3$ and PrAlO$_3$/SrTiO$_3$ interfaces, $\mu$ increases initially with decreasing temperature and drops dramatically for temperatures below 40 K. On the other hand interfaces with LaAlO$_3$ and NdGaO$_3$ show an increase in $\mu$ and tend to saturate at low temperatures. Interesting observation is that even though the $n_s$ (2 K) is nearly the same ($\sim2\times10^{13}$ cm$^{-2}$) for all the interfaces, $R_s$ and $\mu$ exhibit significant divergence, implying that the polar layer has a significant influence on the charge carriers at the polar/nonpolar interface.

Despite these polar layers appear similar in nature, yet there are few differences. Firstly, these polar layers comprise of different chemical elements at A-site (RE cation) and B-site in the perovskite (ABO$_3$) structure. When the polar layer (REBO$_3$) is deposited on top of the TiO$_2$ terminated SrTiO$_3$ (100), the first unit cell at the interface on either side can be viewed as a stack of REO/TiO$_2$/SrO layers. In this picture, across the interface TiO$_2$ layer will have a different electronic environment in presence of different RE cations of polar layer. Secondly, different polar layers cause different lattice mismatch with SrTiO$_3$ which in turn create a different interfacial strain. To further understand these effects, we plotted the A-site cationic dependence of the $n_s$ for corresponding interfaces as shown in Fig. 4(b). Clearly $n_s$ (300 K) are high in the case of La based interface and lower for Nd based interfaces, similar to the case of SrTiO$_3$/REO/SrTiO$_3$ heterostructures reported [20]. However, $n_s$ at 2 K have rather small dependency on the RE cation with $n_s$ is of the order of $1-2\times10^{13}$ cm$^{-2}$ for all interfaces. The RE cation dependency of $n_s$ marks the presence of strong electronic correlation of interface with polar layers.

Now we comment on mobility. Referring back to Fig. 3(c), the $\mu$ drop was only observed in the case of NdAlO$_3$/SrTiO$_3$ and not in the NdGaO$_3$/SrTiO$_3$ (Nd is A-site cation for both cases), implying that we cannot attribute this $\mu$ drop to the proximity effects of RE ions at the interface. So the only other difference these polar layers offer is the lattice mismatch at the interface with the SrTiO$_3$ substrate. Lattice mismatch here is estimated by considering in-plane lattice parameters (a schematic diagram is shown in Fig. 4(a)), among them the NdGaO$_3$/SrTiO$_3$ interface offers the least lattice mismatch ($\sim1.6\%$) and NdAlO$_3$/SrTiO$_3$ the largest ($\sim3.6\%$). Figure 4(c) shows the $\mu$ variation with lattice mismatch; here $\mu$ (300 K) shows modest dependency to the lattice mismatch with values typically of the order of $1-5$ cm$^2$V$^{-1}$s$^{-1}$, which suggests the predominant electron-phonon scattering at 300 K. In contrast, $\mu$ (2 K) shows strong dependence on lattice mismatch: $\mu$ is $\sim500$ cm$^2$V$^{-1}$s$^{-1}$ for NdGaO$_3$/SrTiO$_3$ and only 4 cm$^2$V$^{-1}$s$^{-1}$ for NdAlO$_3$/SrTiO$_3$ (a drop of a factor of 100). It appears that the large lattice mismatch is limiting the $\mu$. We note here that even though the carrier density ($n_s$) at 2 K is almost the same ($2\times10^{13}$  cm$^{-2}$) for all interfaces but a large variation in $\mu$ is observed, indicating strong electronic correlations among carriers. Thus, we argue that the combined effects of strain and strong electron correlation are crucial in controlling the $\mu$. Further, because of a smaller lattice mismatch a higher $\mu$ is expected in the case of NdGaO$_3$ ($\sim1.6\%$) compared to the case of LaAlO$_3$ (mismatch $\sim2.3\%$) due to lesser mismatch, however the mobility values are just comparable in both cases. We attribute this to the residual impurity scattering in SrTiO$_3$ itself.

\begin{figure}
\includegraphics[width=3.4in]{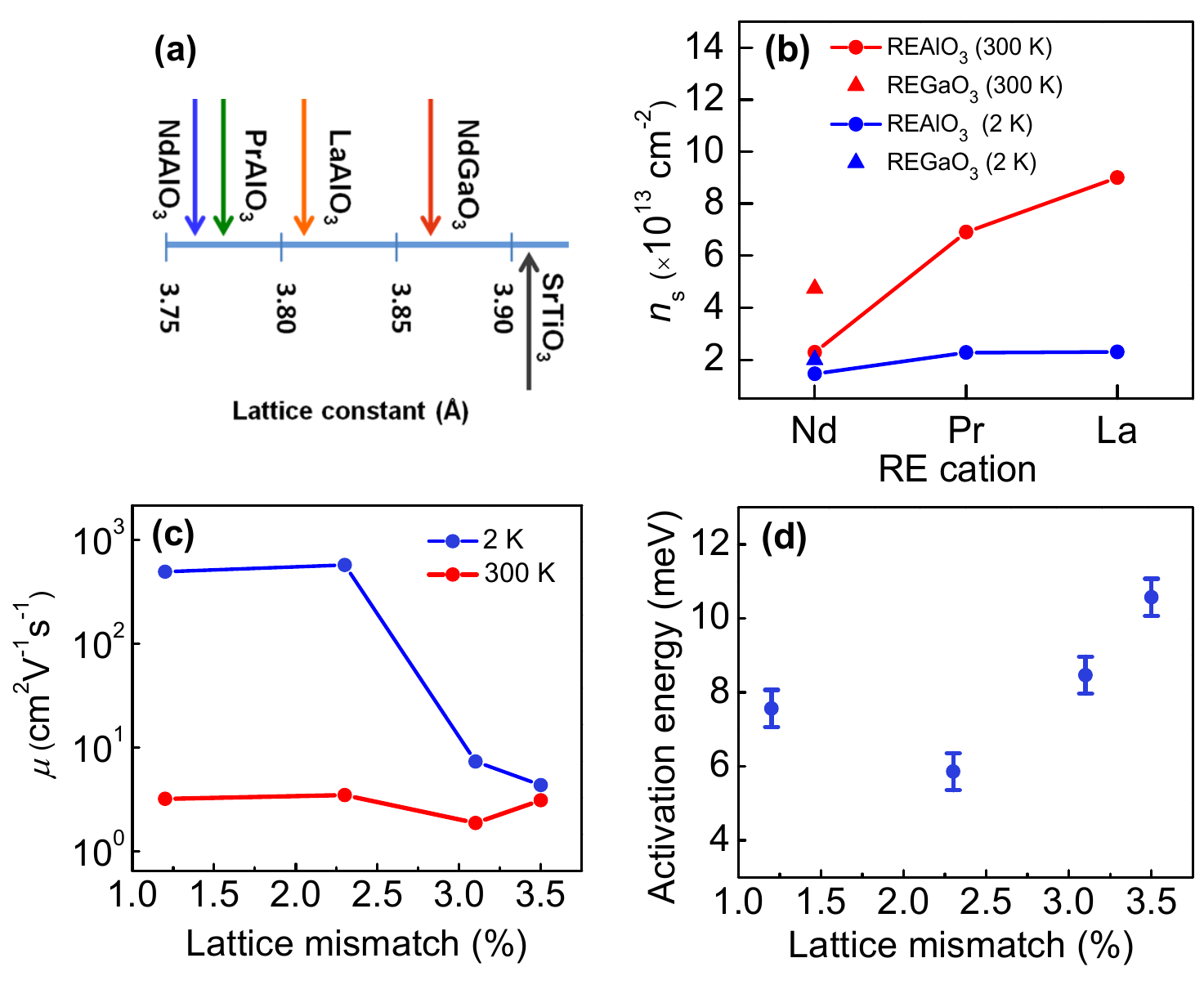}
\caption{\label{fig4} (a) Schematic diagram representing lattice parameters of different polar oxides and SrTiO$_3$ is also shown. (b) $n_s$ dependency on RE cation of polar layer at 300 K and 2 K. (c) $\mu$\ of different interfaces under lattice mismatch at 300 and 2 K. (d) Variation in activation energy as a function of lattice mismatch. }
\end{figure}

As mentioned earlier a large carrier freeze out is observed for the carriers with temperature. To examine this, we extracted activation energy from Arrhenius plots (between 20 and 300 K) of the $n_s$ versus temperature graphs in Fig. 3(b). Activation energies of the order of few meV are in agreement with the previous report on LaAlO$_3$/SrTiO$_3$ [19]. Figure 4(d) shows activation energy as a function of lattice mismatch. Interestingly, activation energy gradually increases with lattice mismatch and it is largest (16 meV) for the NdAlO$_3$/SrTiO$_3$. In general, activation energy indicates the relative position of carrier donor level from the conduction band in energy scale. Its variation with lattice mismatch may suggest relative shift in position of donor level among these interfaces. Recently, it has been proposed that strain could leads to distortion of Ti octahedra through octahedral rotation/tilt in SrTiO$_3$ based heterostructures [20,21], which in turn could alter the position of energy levels. Since electronic effects are quite sensitive to the degree of octahedral distortions, it is further interesting to study the effects of octahedral distortions on novel phases at these interfaces.

\section{Conclusion}

In conclusion, we demonstrate the observation of 2DEG at the NdAlO$_3$/SrTiO$_3$, PrAlO$_3$/SrTiO$_3$ and NdGaO$_3$/SrTiO$_3$ interfaces and suggest that polar discontinuity could be the prime origin of conductivity at these polar/nonpolar interfaces. We showed that the combined effects of strong correlations and interface strain offered by polar layers predominantly control the carrier density and mobility of the 2DEG. Moreover the presence of large octahedral distortions due to interface strain may play an important role in manipulating the novel phases at the interface. Our observations further emphasize the key role of polar layers in these heterostructures and may provide an opportunity to tune the properties at selected polar/nonpolar oxide interfaces.

\begin{acknowledgments}
We thank the National Research Foundation (NRF) Singapore under the
Competitive Research Program (CRP) ``Tailoring Oxide Electronics by
Atomic Control'' NRF2008NRF-CRP002-024, National University of
Singapore (NUS) cross-faculty grant and FRC (ARF Grant No.
R-144-000-278-112) for financial support.
\end{acknowledgments}


\end{document}